\documentstyle[preprint,aps]{revtex}
\begin{document}
\draft
\title{Direct Observation of Field-Induced Incommensurate Fluctuations in
a One-Dimensional S=1/2 Antiferromagnet}
\author{D. C. Dender$^1$, P. R. Hammar$^1$,  Daniel H. Reich$^1$, 
C. Broholm$^{1,2}$, and G. Aeppli$^3$}
\address{
$^1$Department of Physics and Astronomy, The Johns Hopkins
University, Baltimore, Maryland 21218\\
$^2$National Institute of Standards and Technology, Gaithersburg, Maryland 
20899\\
$^3$NEC Research Institute, 4 Independence Way, Princeton, New Jersey 
08540}
\date{\today}
\maketitle
\begin{abstract}
Neutron scattering from copper benzoate, Cu(C$_6$D$_5$COO)$_2
\cdot$3D$_2$O, provides the first direct experimental evidence for 
field-dependent incommensurate low energy modes in  a one-dimensional
spin S = 1/2 antiferromagnet. Soft modes occur for wavevectors
$\tilde{q}=\pi\pm\delta \tilde{q} (H)$ where $\delta \tilde{q} (H) 
\approx 2\pi
M(H)/g\mu_B$ as predicted by Bethe ansatz and spinon descriptions of
the S = 1/2 chain. Unexpected was  a field-induced
energy gap $\Delta (H) \propto H^\alpha$, where $\alpha = 0.65(3)$
as determined from specific heat
measurements.  At $H = 7 $ T  ($g\mu_B H /J = 0.52$), the
magnitude of the gap  varies from $0.06 - 0.3 J$ depending on the
orientation of the applied field.
\end{abstract}


\narrowtext
When a conventional antiferromagnet is subject to a magnetic field the
antiferromagnetically aligned spins reorient perpendicular to the field
in a so-called spin flop transition, and  magnetization
develops through homogeneous canting of the spins in the direction of the
field. A different scenario is expected  for a spin S = 1/2
one-dimensional antiferromagnet 
 \cite{Pytte74,Ishimura77,Muller81},
where the magnetization is
associated with defects in the 
cooperative
spin-singlet ground state.
The effect is accounted for   by mapping the spin
chain  to a one-dimensional system of interacting  fermions \cite{RVB},
popularly referred to as spinons.
The magnetic field causes a Zeeman splitting of the 
half-filled, doubly degenerate spinon band [Fig.~\ref{TheoryFig}(a)], and
the  spin chain
develops soft modes [Fig.~\ref{TheoryFig}(b)] 
at the incommensurate wavevectors which connect
the field-dependent Fermi points. The new periodicity,
given to lowest order by the ratio of the magnetic field to
the spinon velocity, coincides with
the average separation between the magnetization-carrying defects.
While incommensurate spin correlations do not occur in classical
one-dimensional spin systems such as TMMC \cite{Regnault82}, they may 
be a general feature of one-dimensional quantum spin
chains \cite{Muller81,Chitra_Giamarchi}. 
In this letter we present the first direct experimental evidence
for 
incommensurate spin fluctuations in  a uniform antiferromagnetic
S = 1/2 chain.

To search for incommensurate spin correlations, we 
used inelastic neutron scattering to probe 
the wavevector-dependent spin susceptibility of
the model one-dimensional S = 1/2 antiferromagnet (AFM)
copper benzoate, 
$\rm Cu(C_6D_5COO)_2\cdot 3D_2O$.  
In contrast to other systems that have been studied in this context
\cite{Heilmann78,Coldea}, 
in copper benzoate the exchange constant $J$ 
is small enough to permit the large values of the reduced field,
$h=g\mu_B H/J \approx 1$ needed to observe this
effect, while still being large enough to allow the interesting low
energy part of the spectrum
to be explored with a cold neutron triple-axis  spectrometer 
\cite{Dender96}.  
Our most important result is that a magnetic field induces 
new low-energy modes
in the excitation spectrum of copper benzoate at incommensurate 
wavevectors $\tilde{q}=\pi \pm \delta \tilde{q}$, 
where $\delta \tilde{q} (H) 
\approx 2\pi
M(H)/g\mu_B$, with $M(H)$ being the magnetization per spin, 
as predicted by theory \cite{Pytte74,Ishimura77,Muller81}. 
The modes are not completely soft, however, because
the field also induces a gap in the
excitation spectrum both at the incommensurate wave vector and at
$\tilde{q} = \pi$.

First identified as a linear chain AFM by Date {\em et
al.} \cite{Date70}, copper benzoate is centered monoclinic, space group
$I2/c$, with  room temperature lattice constants a = 6.98 \AA, b =
34.12 \AA, c = 6.30 \AA, and $\beta = 89.5^{\circ}$ \cite{Koizumi63}.
Copper ions within a spin chain are separated by ${\bf c}/2$ and
coordinated by edge sharing, tetragonally distorted oxygen octahedra.
The near-neighbor intrachain exchange interaction is
$J = 1.57$ meV \cite{Dender96}.
For this work, deuterated single crystals
were prepared as described previously \cite{Dender96}.  Specific heat
measurements in magnetic fields up to $H = 8.8$ T were made on
individual single crystals of typical mass 0.01 g using the relaxation
method.  For the neutron scattering measurements,
approximately 500 crystals with total 
mass 3.82 g were mutually aligned
to within  5$^\circ$ in the
horizontal $(h0l)$ scattering plane.  Neutron scattering measurements
in fields up to $H = 7$ T $\parallel {\bf \hat{b}}$ were performed on the
SPINS cold neutron triple-axis spectrometer at NIST. 
FWHM beam divergences were 
50$^{\prime}/k_i$(\AA$^{-1})-80^{\prime}-80^{\prime}-215^{\prime}$,
and the fixed final energy was  2.5 meV,  yielding
energy  and wavevector resolution
$\Delta E =
69~\mu$eV and $\Delta \tilde{q}/\pi =  0.02$, 
respectively.
We refer to wave vector transfer along the chain as
$\tilde{q} = {\bf Q}\cdot {\bf c}/2 = l\pi$.  The normalized
magnetic scattering intensity $\tilde{I}(\tilde{q},\omega)$
\cite{Dender96}, was derived from the 
detector count rate by
subtracting $T = 25$ K data as a background, dividing by the
squared magnetic form factor for copper, and
normalizing to incoherent elastic
scattering from 
vanadium.

Figure ~\ref{QScansFig} shows  the $\tilde{q}$-dependence of the  low
energy ($\hbar\omega = 0.21$ meV) and low temperature ($T=0.3$ K)
inelastic magnetic  neutron scattering intensity from copper benzoate
for four  values of $H$. 
At $H = 0$ there is a single peak  centered at $\tilde{q} = \pi$.
This peak is much broader than the instrumental resolution
$\Delta\tilde{q}$, and arises when  $\hbar\omega$ and
$\tilde{q}$ lie within the two-spinon continuum of the S = 1/2 AFM chain
\cite{Muller81,Dender96,BougKar,Nagler}.
In finite field, the
data show additional peaks at incommensurate values of 
wave vector transfer, $\pi\pm\delta\tilde{q}$, 
where $\delta\tilde{q}$ increases with $H$. The data of
Fig.~\ref{QScansFig}
represent the first direct experimental observation of a field-induced 
incommensurate length scale in a uniform spin--1/2 chain and is the 
central result of this paper.

To extract the field dependence of the wavevector $\delta\tilde{q}$
for quantitative comparison to theory, one needs a  specification
of the full $\tilde{q}$- and $\hbar\omega$-dependent magnetic density
of states.  Toward this end, we have performed both specific heat measurements,
which are sensitive to the momentum space average of the density of states,
and constant-$\tilde{q}$ neutron scattering scans at
$\tilde{q}=\pi$ and at $\tilde{q}=1.12 \pi$,
which is the wavevector at which the incommensurate peak 
occurs at $H = 7 $ T.
Figure~\ref{EScansFig} shows the resulting neutron scattering data.
In zero field 
at $\tilde{q}=\pi$ [Fig.~\ref{EScansFig}(a)], 
the threshold for magnetic scattering is $\hbar\omega \approx 0$,
and the 
intensity
decreases monotonically for $\hbar\omega > 0$.
At $\tilde{q}= 1.12 \pi$
[Fig.~\ref{EScansFig}(b)], the threshold has
increased to 
$\hbar\omega \approx 0.8$ meV, with no magnetic scattering
visible below that energy. The lines through the data in the bottom two
frames 
of Fig.~\ref{EScansFig}
were calculated by convolving the 
zero-field
dynamic spin correlation
function derived by Schulz \cite{Schulz}
with the instrumental resolution.

The spectrum changes dramatically for $H = 7$ T, as shown in
Fig.~\ref{EScansFig}(c) and Fig.~\ref{EScansFig}(d).
At $\tilde{q}=\pi$, there is no magnetic scattering
for $\hbar\omega < 0.1$ meV, indicating that a gap has developed in
the spectrum.  Above this gap, a
sharp, resolution-limited mode peaked at
$\hbar\omega = 0.17$ meV 
now marks the onset of the continuum.  
A second, resolution-limited 
mode
also appears at $\tilde{q} = \pi$
at an energy close to the Zeeman energy $g\mu_B H = 0.81$ meV.  
Figure \ref{EScansFig}(b)
reveals that at $H = 7 $ T, the spectrum at the incommensurate wave vector 
$\tilde{q}= 1.12 \pi$ also has a gap to a resonant mode 
at $\hbar\omega = $ 0.22 meV.
The dependence of the intensity of the resonant modes 
on the angle between the total
wavevector transfer ${\bf Q}$ and the chain axis
is consistent with the theoretical expectation \cite{Muller81} 
that the
incommensurate mode is polarized parallel to the field, and the mode at
$\tilde{q} = \pi$ is  polarized perpendicular to it \cite{DenderToBe2}.

We now turn to the specific heat measurements, which explore the 
field dependence of the gap.
 Figure ~\ref{Cp} shows  specific heat data for
copper benzoate for the same values and orientation
 of the applied field ($H \parallel {\bf \hat{b}}$)
as for the data in Fig.~\ref{QScansFig}. 
The total specific heat for $H = 0$ and 3.5 T is shown in the inset, plotted
as $C_{tot} /T$.  
A  small lattice  
contribution to $C_{tot}$ below $T = 1$ K is visible in the slight
temperature dependence of $C_{tot} /T$ for $H = 0$.  
Fitting the $T < 1$ K zero-field data to $C_{tot} /T = a + bT^2$ gives
$b = 0.024(5) $ J/mol~K$^4$.  This lattice term was subtracted
from all the data, and 
the main part of Fig.~\ref{Cp} shows 
the resulting magnetic specific heat $C$.
The linear  magnetic specific heat we observe in zero field is
consistent with the low-energy linear dispersion relation
of the spinons in
the 1D spin--1/2 AFM.  
The fit shown gives $C(T) = 0.68(1) R (k_B T/J)$,
in excellent
agreement with the  theoretical value of
$C(T) = 0.7 R (k_B T/J)$ \cite{Bonner_Fisher64}. The data set an 
upper limit of $\approx$  9 $\mu$eV for any zero-field gap and an upper limit
$\Delta S < 3\times 10^{-3}R\ln 2$ on the entropy change 
of any magnetic phase transition
for $T>0.1$ K \cite{Takeda80}.

In finite fields and at low $T$, $C$ is suppressed below its zero field value.
As $T$ increases, $C(H,T)$ rises above the
zero-field curve, before settling back down to it at high $T$ 
(inset to Fig.~\ref{Cp}).
This  behavior indicates a  transfer of  spectral weight from low to higher
energies,
and is a clear signature of a field-induced gap $\Delta(H)$.

To determine the field dependence of   $\Delta$, we fit the
$H \ne 0$ data in Fig.~\ref{Cp} to 
\begin{equation}
C = \frac{\tilde{n} R}{\sqrt{2\pi}} \left (\frac{ \Delta}{k_B T}\right)^{3/2}
     \frac{\Delta}{v}\exp(-\Delta/k_B T) ,
\label{CpEqn}
\end{equation}
which is the low--$T$ specific heat
\cite{Troyer94} for $\tilde{n}$ species of non-interacting
one-dimensional bosons with a gap and dispersion relation 
\begin{equation}
\hbar\omega (\tilde{q})=\sqrt{\Delta^2+(v(\tilde{q}-\tilde{q}_0))^2},
\label{DispRel}
\end{equation}
A term  proportional to $H^2/T^2$ was included in the fit
to take  into account the small nuclear spin contribution.
The  fits are shown as solid lines in Fig.~\ref{Cp},
 and Fig. \ref{Hdependence}(b)  
summarizes the field dependence of the gaps derived from this analysis
as well as for data we have taken with $H$ along the other principal magnetic
directions, $\bf c^{\prime\prime}$ and $\bf a^{\prime\prime}$
\cite{cubspin}.
The average of the gaps measured by neutron scattering for $H=7$ T 
$\parallel {\bf \hat{b}}$, at $\tilde{q}=\pi$ and 
1.12$\pi$ (filled symbols in Fig. ~\ref{Hdependence}(b)) correspond nicely
to the gap derived from specific heat.  Moreover the values of 
$v/\tilde{n}$ = 0.49(3), 0.55(3), and 0.96(5) meV/mode for
$H \parallel$ $\bf b, c^{\prime\prime}$ and $\bf a^{\prime\prime}$, 
respectively, are not far from the crude estimate
$v/\tilde{n} \approx 0.5\pi J/6 = 0.41$ meV/mode, where 
$\tilde{n}=6$ is the number of soft modes in a Brillouin zone.
The field 
dependence of the gaps is described by the
power-law  $\Delta (H) = A H^\alpha$
with $\alpha = 0.65(3)$.  The prefactors $A$ are in
ratios 1~:~2.0~:~0.55 for $H$ 
applied along the $\bf b, c^{\prime\prime}$ and $\bf a^{\prime\prime}$
directions, respectively. A slightly better fit to the
$\bf c^{\prime\prime}$
data is obtained by including a finite critical field:
$\Delta (H) = A (H-H_c)^\alpha$, where $H_c=0.25(3)$ T and 
$\alpha = 0.58(3)$.
We note that small field-dependent gaps are not found in the 
classical easy plane AFM
spin chain TMMC, where a field in the easy plane induces a central
diffusive soliton mode and a gap $\Delta (H)=g\mu_B
H$ \cite{Regnault82,Heilmann81}.

With the information about
the field-dependence of the gap derived from the specific heat data,
we can  extract the field-dependent incommensurate
wave vector from the neutron scattering
data in Fig.~\ref{QScansFig}. 
We approximate the low
energy response as  resonant modes with dispersion relations given 
by Eq. (\ref{DispRel}),
where $\tilde{q}_0=\pi$ or $\pi\pm\delta\tilde{q}(H)$
(see Fig.~\ref{TheoryFig}(b)). The velocity
$v$ was fixed at the zero field value $\pi J /2 $ 
\cite{Cloizeaux62}, and the gap
$\Delta(H)$ was taken from the specific heat measurements.  
Although it neglects the continuum scattering above the
resonant modes that is visible in Figs.~\ref{EScansFig}(a) and 
\ref{EScansFig}(b), this model
provides a reasonable fit for the finite field constant--$\hbar\omega$ scans
in Fig.~\ref{QScansFig}. 
Note that for $H = 3.5$ T, it appears that
a single branch of the dispersion relation at the
incommensurate wave vector is resolved, while the other merges into a
broad central peak at $\tilde{q}=\pi$. Thus the incommensurate wavevector
at this field actually lies at the local minimum between the peaks. 
This is in contrast to the $H = 5$ T and $H = 7$ T data, where
the incommensurate modes are well-separated from the 
$\tilde{q} = \pi$ mode, and the energy of the scan is very close to the gap.
The fits establish that the
incommensurate peaks are resolution-limited, and 
also provide  a systematic
way of deriving the field-dependent incommensurate wavevector
$\delta\tilde{q}(H)$.  This is shown as a
function of $H$ in Fig.~\ref{Hdependence}(a). The solid line in this
figure is the theoretical prediction
\cite{Muller81}:  $\delta\tilde{q}(H)=2\pi M(H)/g_b\mu_B$
with $g_b = 2.06$ \cite{cubspin}.
$M(H)$ is 
calculated using Eq. (2.24) of Ref.~\cite{Muller81}, which
provides an excellent description of the magnetization
for copper benzoate \cite{Yosida83}. 
Despite the complications associated
with the field-induced  gaps, the theory
accounts well for the field-dependent positions of the
incommensurate soft modes. 

In summary we have provided the first direct experimental evidence for 
field-dependent incommensurate soft modes in a S=1/2 one dimensional
antiferromagnet.
Thus, one of the key features of the spinon description of 
S=1/2 AFM chains, namely a chemical potential which can be
regulated by an external field, has finally been verified.
Our results also have implications beyond the nearest-neighbor
coupled S=1/2 chain. For example, theoretical studies predict 
that  both long-range coupled models 
\cite{Haldane_Talstra} and 
systems such as ladders and dimerized chains with intrinsic gaps
\cite{Chitra_Giamarchi}
should
show similar behavior, indicating that field-induced 
incommensurabilities are a general property of quantum
spin chains.  The incommensurate 
lattice distortions recently observed  
in TTF-CuBDT \cite{Kiryukhin95} and  CuGeO$_3$ 
\cite{Kiryukhin96}
may therefore be closely related to our results.
In copper benzoate, the field also introduces gaps in the excitation
spectrum both at the commensurate and at the incommensurate wave
vector, something which was not predicted by theories for 
isotropic Heisenberg
spin chains.  Possible causes for the gaps 
are diagonal as well as off-diagonal exchange anisotropies, and a
staggered g-factor anisotropy \cite{cubspin} 
which induces a small effective staggered
field when a homogeneous external field is applied .

We are grateful for discussions with A. Millis,
and for the hospitality of K. Clausen at
Ris\o\ National Laboratory, where preliminary data were collected.
NSF grants 
DMR-9302065, and DMR-9453362 supported work at JHU. 
This work utilized neutron research facilities supported by NIST and the
NSF under Agreement No. DMR-9423101. 
DHR acknowledges support from the David and Lucile Packard Foundation.

\begin{figure}
\caption{(a) Schematic of the Zeeman-split bands of the fermion
mean field theory for the S=1/2 AFM chain in a magnetic field $H$.
The Fermi energy is at $E = 0$.
Solid (dashed) arrows show
incommensurate (commensurate) spanning vectors. (b) Boundaries
of the resulting spinon continuua, for spin fluctuations parallel
(solid lines) and perpendicular (dashed lines) to $H$.
Note incommensurate soft modes near $\tilde{q} = \pi$.
Horizontal dotted line shows trajectory of scans in Fig.~2.}
\label{TheoryFig}
\end{figure}

\begin{figure}
\caption{
Magnetic scattering at $\hbar\omega = 0.21$ meV along the (0.3,0,$l$)
 directionfor four values of  
magnetic field $H$ at $T = 0.3$ K. The solid lines for $H \ne 0$
are fits to resolution-limited
spin wave scattering as described in the text. The line through 
the $H=0$ T data is the theory  \protect\cite{Schulz} for the 
dynamic spin correlation function convolved with the
experimental resolution. A correction factor varying from 0.7
to 0.95 over the range of the scan was applied to the model to take into account
the measured $\tilde{q}$ dependent sample transmission.  }
\label{QScansFig}
\end{figure}

\begin{figure}
\caption{Energy dependence of the magnetic scattering
intensity at $T = 0.3$ K for
${\bf Q}=(0.3,0,1)$ and ${\bf Q}=(0.3,0,1.12)$ at $T=0.3$ K. 
The latter wave vector corresponds to
the position of the incommensurate maxima in the $H=7$ T
constant--$\hbar\omega =0.21$ meV scan.  The
solid lines through the $H=0$ T data are the theoretical dynamic
correlation function \protect\cite{Schulz} convolved with the
experimental resolution. The dashed lines are
the average of the data in (b) for 0.08 meV $<\hbar\omega<0.6$ meV, which
is a good measure of the over-subtraction caused by isotropic
magnetic scattering contained in the $T=25$ K data
used as a background.  
}
\label{EScansFig}
\end{figure}

\begin{figure}
\caption{Specific heat of copper benzoate as a function of temperature for four
values of applied field $H \parallel {\bf \hat{b}}$.
Inset: total specific heat plotted as $C_{tot}/T$ for $H = 0$ and 3.5 T.
Main panel: magnetic heat capacity $C$ after subtraction of phonon 
contribution.
Solid lines are from fits  described in the text.}
\label{Cp}
\end{figure}

\begin{figure}
\caption{
(a) Field dependence of the displacement $\delta\tilde{q}$  of the 
incommensurate side peaks from $\tilde{q}=\pi$ in copper benzoate, 
as derived from fits to the data shown in Fig.~2.  The solid line
is the theoretical curve from Ref.~\protect\cite{Muller81}.
(b) Field dependence of the energy gap derived
from fits to specific heat data such as those shown in Fig.~4. Data
for fields along the three principal magnetic directions
are shown. Filled symbols are the gaps measured at $\tilde{q}=\pi$
and $\tilde{q}=1.12\pi$ by neutron scattering. The solid lines are 
from fits to
power-laws  described in the text.
}
\label{Hdependence}
\end{figure}

\end{document}